# Donnan Equilibrium Revisited:
# Coupling between Ion Concentrations, Osmotic Pressure, and Donnan Potential

Gang Chen[1]

Department of Mechanical Engineering
Massachusetts Institute of Technology
Cambridge, MA 02139

**Abstract**

This paper discusses the little-known fact that Donnan's equilibrium criteria established over 100 years ago neglected the coupling between ion concentrations and the osmotic pressure. Such coupling can be treated based on general thermodynamic considerations including the solvent equilibrium, leading to a membrane potential that consists of not only the classical Donnan potential term but also an additional term due to the osmotic pressure, and the existence of a membrane potential even when the impermeable species are not charged. This coupled treatment is in conflict with the well-established Poisson-Boltzmann equation and Nernst-Planck equation, but is consistent with the extension of these equations including the solvent effects by Freise and Schlogl, enables us to view the electrical double layer equilibrium as Donnan equilibrium.

[1] Email: gchen2@mit.edu

**Introduction**

Donnan equilibrium (also called Gibbs-Donnan equilibrium) refers to the imbalance of ion species across two solutions separated by an interface that prevents at least one species to pass through [Donnan, 1924]. The interface could be a semipermeable membrane, or between a polyelectrolyte polymer gel and its surrounding solution [Brannon-Peppas and Peppas, 1991], or between two different polymer gels [Ohki, 1965; Guo et al., 2016]. There are two key phenomena that are often associated with the Donnan equilibrium: osmotic pressure and Donnan potential. The former is a pressure difference across the membrane or interface and the latter is an electrical potential difference between the two liquids. These two phenomena are observed and exploited in a wide range of systems and technologies, from biology [Sperelakis, 2001; McLaughlin, 1989] to desalination [Mohammad et al., 2015] to batteries and fuel cells [Knehr and Kumbur, 2011].

Donnan established relationships between ion concentrations in the two regions at equilibrium [Donnan, 1924; 1955], which can be combined with the charge neutrality requirement to solve for the concentrations of the ions in each region. Once the concentrations are known, one can easily calculate the osmotic pressure from van't Hoff's law and the Donnan potential from the ion concentration ratio. Donnan equilibrium criteria are widely used in different fields such as cell biology [Sperelakis, 2001; McLaughlin, 1989], membrane desalination [Greenlee et al., 2009], and polyelectrolyte hydrogel expansion [Flory, 1953].

Donnan's equilibrium criteria are a one-way street. One first determines the concentrations of ions, and then uses the concentrations to compute the osmotic pressure and the Donnan potential. The latter two are not interrelated, i.e., they do not influence each other. Donnan derived the equilibrium relations by requiring that the chemical potential of each mobile species in the two compartments equaling each other. Since Donnan's work, there have been numerous studies rederiving Donnan's criteria [Philipse and Vrij, 2011], most of them led to the same results as Donnan originally obtained. The exceptions, to the best of the author's knowledge, were the work of Morales and Shock [1941] and Scotto et al. [2016], neither had drawn much attention. However, a recent review [Fahlman et al., 2019] stated "Scotto et al.'s work in 2016 is the only correct complete theory on the Donnan equilibrium."

In this work, we will start from basic thermodynamic relations to explain what is missing in Donnan's original analysis, and derive corrected membrane equilibrium criteria, consistent with Refs. [Morales and Shock, 1941; and Scotto et al., 2016], using four example problems. These examples show that the concentrations of ions across a membrane are coupled to the osmotic pressure; and that the Donnan potential consists of not only the classical Donnan's potential term due to the ion concentration difference, but also an additional term from the osmotic pressure. This analysis also predicts the existence of a membrane potential even if the impermeable species are not charged, which can be understood as the generation of ion separation and membrane potential by osmotic pressure. We will discuss the connection of the Donnan equilibrium to electrical double layer and ion transport across membranes. While the classical Donnan criteria are consistent with the Poisson-Boltzmann and the Nernst-Planck equations, the coupled equilibrium criteria are consistent with the equations derived by Freise (1952) and Schlogl (1966) that address shortcomings in aforementioned equations. Recognizing the couplings discussed in this paper calls for reconsideration of many phenomena considered in the past in literature.

**Donnan's Approach vs. Coupling Approach**

Let us consider a solution consists of $\chi_a, ..\chi_i, ..\chi_n$ and $\chi_s$ mole fractions of components a,..., i,..., n, and s, where s is used specifically for the solvent such that

$$\chi_a + ... + \chi_i + ... + \chi_n + \chi_s = 1 \tag{1}$$

We use $c_a$, $c_b$, ...$c_i$, ..., $c_n$, $c_s$ to represent their corresponding molar concentration (mole per liter). The chemical potential of ith component on a molar basis can be written as [Lewis et al., 1961; DeVoe, 2020; Ellgen]

$$\mu_i(T, p, \varphi, \chi_a, ..\chi_i, ..\chi_n, \chi_s) = \mu_i^o(T, p^o) + \bar{v}_i(p - p^o) + z_i F(\varphi - \varphi^o) + RT\ln(\gamma_i \chi_i) \tag{2}$$

where F is the Faraday constant and R is the universal gas constant, $\bar{v}_i$ the molar volume, T and p are the temperature and pressure of the system, $z_i$ the net valence of the species i (including the sign), and $\gamma_i$ the activity coefficient. $\mu_i^o(T, p^o)$ the chemical potential of standard state (T,$p^o$). Donnan's treatment neglected the pressure term in Eq. (2) [Donnan, 1955]. Overbeek [1956] started from Eq. (2) but argued that the pressure term is too small and can be neglected. However, it is exactly this term that is related to the osmotic pressure. This neglect led Donnan to arrive at the traditional Donnan equilibrium criteria and the decoupling of the osmotic pressure with the electrical potential.

We will first examine an example discussed in Donnan's classical review paper [Donnan, 1924]. He considered a membrane separating two solutions. Initially BCl with concentration $c_1$ existed in compartment 1 and HCl with concentration $c_2$ in compartment 2 (Fig.1) (one can easily show that the final equilibrium state does not depend on if the mobile ions are initially only in 2). He assumed that (1) both BCl and HCl are strong electrolytes and are fully ionized, (2) cation $B^+$ cannot go through the membrane, (3) $H^+$ and $Cl^-$ ions can go through the membrane, (4) the solutions are ideal, i.e., the activity coefficients for all components are 1, and (5) the two compartments have equal volume. Some $H^+$ and $Cl^-$ ions will cross the membrane and the two sides will reach equilibrium. At equilibrium, if the concentration of $H^+$ in 1 is y, the charge neutrality requirement mandates that $Cl^-$ concentration should be $c_1+y$. Equal volume assumption means that the concentrations of $H^+$ and $Cl^-$ in region 2 is reduced by y. Donnan invoked the equal chemical potential on two sides for mobile ions without considering the pressure term in Eq. (2). Applying Eq. (2) to $H^+$ and $Cl^-$ ions in each compartment, setting the chemical potentials of each ion species in the two compartments equaling each other leads to two equations, and further eliminating the electrostatic potential in these two equations, we get

$$\frac{c_{H^+,1}}{c_{H^+,2}} = \frac{c_{Cl^-,2}}{c_{Cl^-,1}} \quad \text{or} \quad \frac{y}{c_2-y} = \frac{c_2-y}{c_1+y}$$

which can be easily solved

$$y = \frac{c_2^2}{c_1+2c_2} \quad \text{or} \quad y^* = \frac{c_2^{*2}}{c_1^*+2c_2^*} \quad (3)$$

where the superscript "*" is used to denote the mole fraction-based quantities ($y^*=y/c_o$, $c_1^*=c_1/c_o$, $c_2^*=c_2/c_o$) with $c_o$ the total solution molar concentration.  Once the ion concentrations in the two compartments are known, the osmotic pressure can be calculated by applying Eq.(2) to the solvent on each compartment and setting their chemical potential equaling each other,

$$\Pi = p_1 - p_2 = \frac{RT}{v_s} \ln \frac{x_{s,2}}{x_{s,1}} \approx \frac{RT}{v_s c_o} \left[ (c_{H^+,1} + c_{Cl^-,1} + c_{B^+,1}) - (c_{H^+,2} + c_{Cl^-,2}) \right] = \frac{2RT}{\bar{v}_s} \frac{c_1^*(c_1^*+c_2^*)}{c_1^*+2c_2^*} \quad (4)$$

where the linearization of the logarithm function after substituting the solvent concentration using Eq.(1) can be justified based on the assumption that the solute concentration is much smaller than one.  Such linearization leads to the familiar van't Hoff law.  And the Donnan potential is

$$\varphi = \varphi_1 - \varphi_2 = -\frac{RT}{F} \ln \frac{c_{H^+,1}}{c_{H^+,2}} = -\frac{RT}{F} \ln \frac{c_2^*}{c_1^*+c_2^*} \quad (5)$$

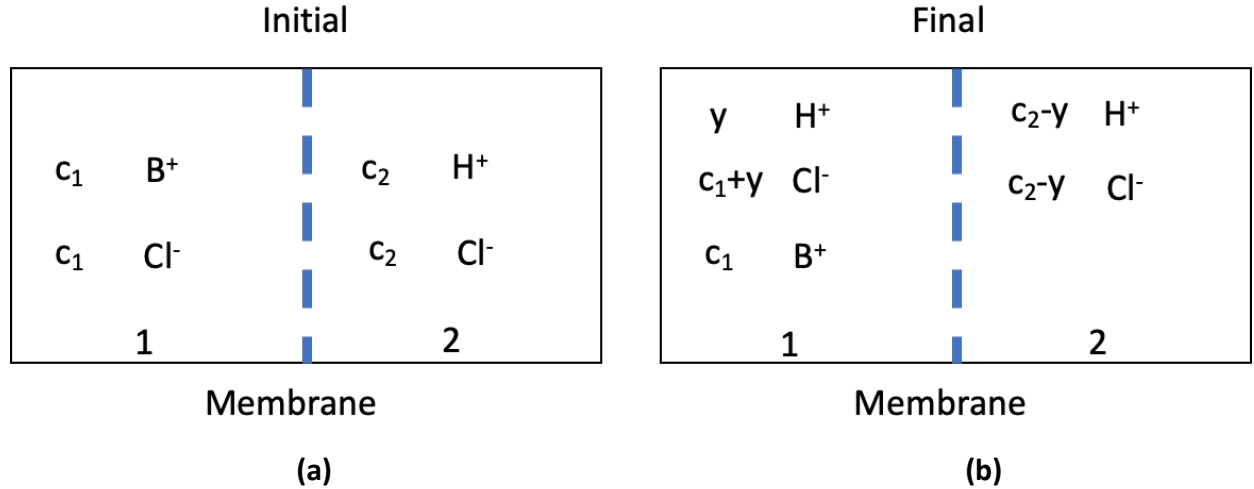

**Figure 1  Membrane equilibrium example 1.** A membrane impermeable to cation $B^+$ separates two strong electrolytes containing BCl and HCl, (a) initial and (b) final state.

Instead of Donnan's approach, we now keep the pressure term in Eq. (2), and apply it to $H^+$ ions in compartments 1 and 2. At equilibrium, there is no net exchange of the $H^+$ ions nor $Cl^-$ ions, which means the chemical potentials on the two sides are equal to each other for each species. This equilibrium leads to,

$$0 = \bar{v}_{H^+}\Pi + F\varphi + RT\ln\frac{\chi_{H^+,1}}{\chi_{H^+,2}} \tag{6}$$

$$0 = \bar{v}_{Cl^-}\Pi - F\varphi + RT\ln\frac{\chi_{Cl^-,1}}{\chi_{Cl^-,2}} \tag{7}$$

Adding Eqs. (6) and (7), we have

$$0 = (\bar{v}_{H^+} + \bar{v}_{Cl^-})\Pi + RT\ln\left(\frac{\chi_{H^+,1}}{\chi_{H^+,2}} \times \frac{\chi_{Cl^-,1}}{\chi_{Cl^-,2}}\right) \tag{8}$$

We can see now that the ion concentrations on the two sides are actually dependent on the osmotic pressure. To solve for the ion concentrations, we need to use the fact that the solvents in the two compartments are also in equilibrium. Applying the condition of equal chemical potential for the solvent on the two sides, we get we get,

$$\bar{v}_s\Pi = -RT\ln\frac{\chi_{s,1}}{\chi_{s,2}} = -RT\ln\frac{1-(\chi_{H^+,1}+\chi_{Cl^-,1}+\chi_{B^+,1})}{1-(\chi_{H^+,2}+\chi_{Cl^-,2})}$$
$$\approx RT[(\chi_{H^+,1} + \chi_{Cl^-,1} + \chi_{B^+,1}) - (\chi_{H^+,2} + \chi_{Cl^-,2})] = 2RT(c_1^* - c_2^* + 2y^*) \tag{9}$$

where again the approximation is due to linearization of the logarithm function. Equation (9) is the same as we wrote done in Eq. (4) except now that y* is an unknown. For the current approach, it is actually better to use the logarithm form for the solvent osmotic pressure and linearize it later. Eliminating the osmotic pressure in Eqs. (8) and (9), we arrive at

$$\frac{\bar{v}_{HCl}}{\bar{v}_s}\ln\frac{\chi_{s,1}}{\chi_{s,2}} = \ln\frac{y^*(c_1^*+y^*)}{(c_2^*-y^*)^2} \tag{10}$$

where $\bar{v}_{HCl} = \bar{v}_{H^+} + \bar{v}_{Cl^-}$ is the molar volume of HCl in the solvent. Equation (10) can be expressed as

$$\left[\frac{1-(2c_1^*+2y^*)}{1-(2c_2^*-2y^*)}\right]^\Gamma = \frac{y^*(c_1^*+y^*)}{(c_2^*-y^*)^2} \tag{11}$$

where $\Gamma = \frac{\bar{v}_{HCl}}{\bar{v}_s}$ is the ratio of the molar volumes. The above equation can be solved for y* for given c$_1$*, c$_2$*, and $\Gamma$. If the solute concentrations are much lower than the solvent concentration, we can linearize the left-hand side of Eq. (11) to obtain

$$1 - 2\Gamma(c_1^* - c_2^* + 2y^*) \approx \frac{y^*(c_1^*+y^*)}{(c_2^*-y^*)^2} \tag{12}$$

The above equation still does not lend to explicit expressions for y* and needs to be solved numerically. Donnan's traditional equilibrium criteria require the left-hand-side of Eqs. (11) and (12) equaling unity, effectively decoupling the solvent from the ionic concentration. The difference of the current approach from the traditional Donnan criteria becomes larger if the left-hand-side deviates from unity.

Once Eq.(11) or (12) is solved, we can compute the osmotic pressure from Eq. (8) or (9), and use either Eq. (6) or (7) to calculate the Donnan potential. Using Eq. (7), the Donnan potential can be written as

$$\varphi = \frac{RT}{F}\left(ln\frac{\chi_{Cl^-,1}}{\chi_{Cl^-,2}} + \frac{\bar{v}_{Cl^-}}{RT}\Pi\right) = \frac{RT}{F}\left(ln\frac{c_1^*+y^*}{c_2^*-y^*} - \frac{\bar{v}_{Cl^-}}{\bar{v}_s}ln\frac{1-(2c_1^*+2y^*)}{1-(2c_2^*-2y^*)}\right) \quad (13)$$

The first term in Eq. (13) is the traditional Donnan potential, $\varphi_c$, arising from the ion concentration difference. The second term, which is not in Donnan's treatment, is due to the osmotic pressure, $\varphi_\Pi$. Since $Cl^-$ concentration in 1 is higher than in 2 (assuming $c_1>c_2$), $Cl^-$ has the tendency to diffuse from 1 into 2. The first term in Eq. (13) is positive, which means $Cl^-$ is attracted back to compartment 1 due to the Donnan potential. Meanwhile, the solvent diffusion creates a higher pressure in compartment 1 (again assuming $c_1>c_2$), which also tends to push $Cl^-$ from 1 to 2. The additional membrane potential $\varphi_\Pi$ is created to balance the osmotic pressure driving force on the ions. This explains the origin of the additional membrane potential term.

In Fig.2(a)-(d), we compare the equilibrium composition, osmotic pressure, membrane potential, and the additional membrane potential obtained from solving Eqs. (11) and (13), which we call "exact". Solution obtained from approximation (12) be called "approximate". We call results based on Eqs. (4) and (5) as "Donnan". For solvent, we took water $\bar{v}_s$=1.8x18$^{-5}$ m$^3$/mol. Although the molar volumes of H$^+$ and Cl$^-$ are generally not equal to each other, we will take $\bar{v}_{H^+} = \bar{v}_{Cl^-} = 1.15 \times 10^{-5}$ m$^3$/mol for now, and discuss the molar volume effect later. In the range of concentration studied, the difference in y between the Donnan approach and the "exact" one presented here can differ as much at 20% in the high concentration range [Fig.2(a)]. For example, at $c_1/c_o$=10.5% and $c_2/c_o$=1%, Donnan formula gives y*=7.99x10$^{-4}$ while Eq.(11) leads to y*=6.29x10$^{-4}$. The difference in the osmotic pressure [Fig.2(b)], however, is negligible between the two approaches. This is because the osmotic pressure is mainly decided by the differences between $c_1$ and $c_2$ [see Eq. (9)], and y is not large enough to impact the osmotic pressure. However, the membrane potential calculated from the traditional Donnan approach and current approach can differ by ~10% [Fig.2(c)]. For example, at $c_1/c_o$=10.5% and $c_2/c_o$=1%, Donnan's formula gives φ=63.2 mV while Eq.(13) gives φ=69.9 mV. In Fig.2(d), we show the contribution to the membrane potential from the osmotic pressure, which is not included in Donnan's approach. In Fig.2(d), we show the contribution to the membrane potential from the osmotic pressure, which is not included in Donnan's approach.

Now turn to the first example Donnan considered in his paper [1924], as sketched in Fig. 3, in which two electrolytes NaA and KA with initial concentrations $c_1$ and $c_2$ are separated by a membrane and the anion A$^-$ cannot across the membrane. Under same assumptions of strong electrolytes, traditional Donnan equilibrium approach gives

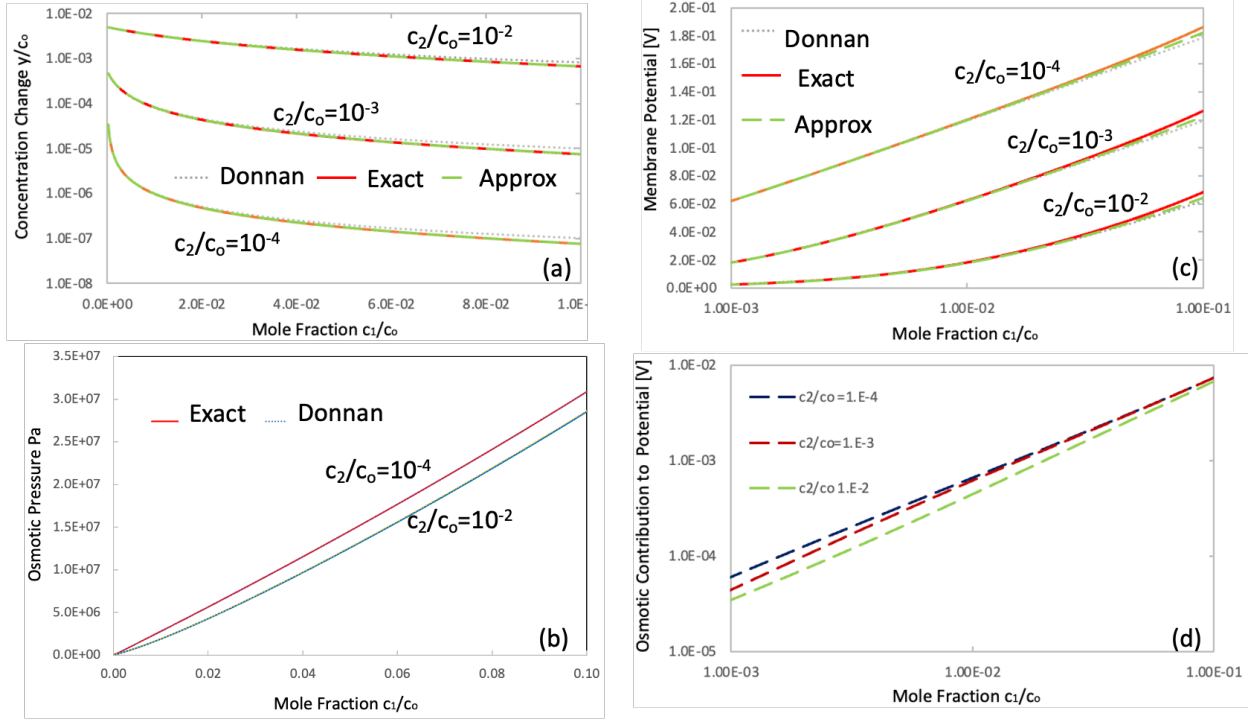

**Figure 2. Comparison between coupled treatment and Donnan's approach.** (a) change in concentration, (b) osmotic pressure, (c) membrane potential, and (d) osmotic contribution to the membrane potential, which is not included in the Donnan approach. "Exact" means from coupled formulation, "Approx" is based on linearizing the osmotic pressure as in Eq. (12), and "Donnan" is based on the Donnan approach.

$$y^* = \frac{c_1^* c_2^*}{c_1^* + c_2^*} \qquad \Pi = \frac{2RT}{\bar{v}_s}(c_1^* - c_2^*) \qquad \varphi = -\frac{RT}{F} \ln \frac{c_1^*}{c_2^*} \qquad (14)$$

To include the pressure effect on the Donnan potential, we follow same procedures described above. For $Na^+$ ions, the chemical potential balance leads to

$$0 = \bar{v}_{Na^+} \Pi + F\varphi + RT \ln \frac{c_{Na^+,1}}{c_{Na^+,2}} \qquad (15)$$

Similarly, for $K_a^+$ ions, we can write

$$0 = \bar{v}_{K^+} \Pi + F\varphi + RT \ln \frac{c_{K,1}}{c_{K^+,2}} \qquad (16)$$

Taking the difference of Eqs. (15) and (16), we have

$$(\bar{v}_{Na^+} - \bar{v}_{K^+})\Pi = -RT \left[ \ln \frac{c_1^* - y^*}{y^*} - \ln \frac{y^*}{c_2^* - y^*} \right] \qquad (17)$$

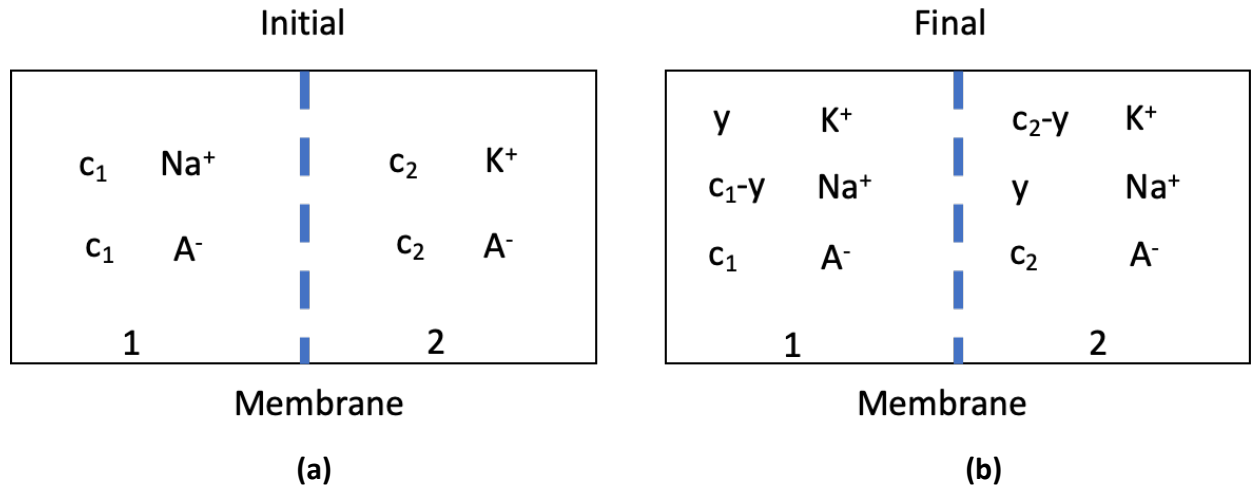

**Figure 3** Donnan equilibrium between two systems originally containing $c_1$ NaA and $c_2$ KA in region 1 and 2 with the membrane impermeable to anion A$^-$, (a) before and (b) after equilibrium.

We obtain the osmotic pressure from the solvent equilibrium,

$$\Pi = -\frac{RT}{\bar{v}_s} \ln \frac{\chi_{s1}}{\chi_{s2}} \approx \frac{2RT}{\bar{v}_s}(c_1^* - c_2^*) \tag{18}$$

If one assumes that the molar volume of Na$^+$ and K$^+$ are same, solutions of Eqs. (15) and (16) for the concentration and hence the osmotic pressure as given by Eq. (18) will be identical to these obtained by Donnan. Even in this case, however, the membrane potential that can be obtained from the same set of equations will be different from that given by Donnan. This potential can be expressed as

$$\varphi \approx -\frac{RT}{F}\left(\ln\frac{c_1^*}{c_2^*} + \frac{2\bar{v}_{K^+}}{\bar{v}_s}(c_1^* - c_2^*)\right) \tag{19}$$

In Eq. (19), the second term is extra that is not included in the Donnan expression. It is contributed by the osmotic pressure. If we assume $c_1 > c_2$, it can be easily seen that both the concentration gradient and the pressure gradient drive K$^+$ ions from 1 to 2, and hence a negative potential gradient develops to resist the combined effects of concentration and pressure differences.

The effect of the molar volume is best seen for a nonsymmetric electrolyte. Taking again an example in Donnan's paper, in which KA in the previous example is replaced by CaA$_2$ (Fig.4). Figure 4b shows the species transferred meeting the charge neutrality requirement. At equilibrium, we have same expression for Na$^+$ ions as in Eq. (15). But for the Ca$_2^+$ ions, applying the chemical potential balance leads to

$$0 = \bar{v}_{Ca^{2+}}\Pi + 2eN_A\varphi + RT\ln\frac{c_{Ca^{2+},1}}{c_{Ca^{2+},2}} \tag{20}$$

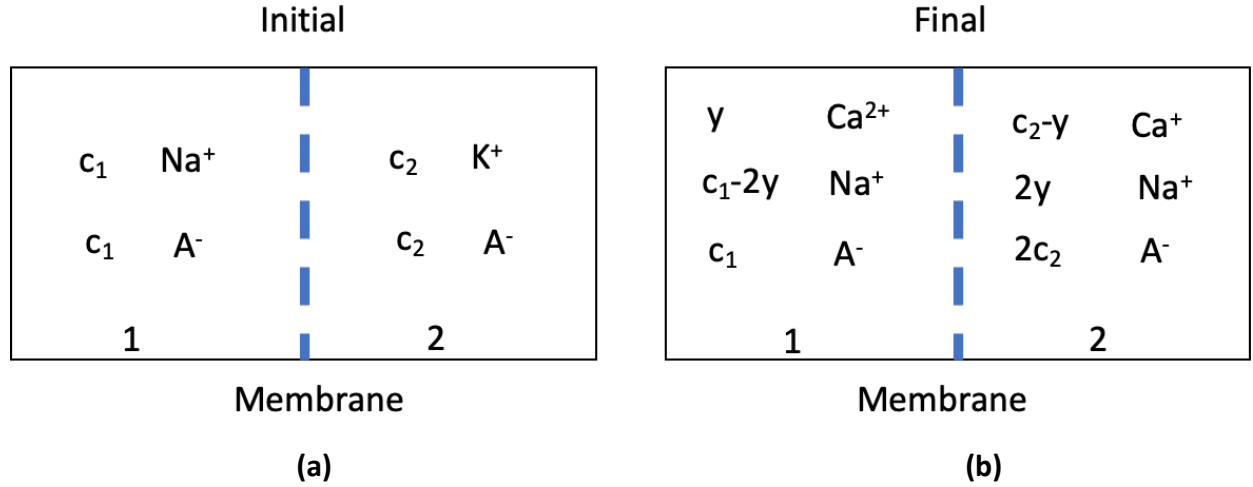

**Figure 4** Example of nonsymmetric electrolyte (a) before and (b) after equilibrium with the membrane impermeable to anion A$^-$.

Eliminating the potential difference from Eqs.(15) and (20), we get,

$$(2\bar{v}_{Na^+} - \bar{v}_{Ca^{2+}})\Pi = -RT\left[\ln\left(\frac{c_1^*-2y^*}{2y^*}\right)^2 - \ln\frac{y^*}{c_2^*-y^*}\right] \tag{21}$$

The molar volumes on the left-hand side clearly do not cancel out. So, the classical Donnan's equilibrium condition cannot be satisfied. To eliminate the osmotic pressure, we use the solvent equilibrium condition, which leads to

$$\left(\frac{1-(2c_1^*-y^*)}{1-(3c_2^*+y^*)}\right)^\Gamma = \left(\frac{c_1^*-2y^*}{2y^*}\right)^2 \frac{c_2^*-y^*}{y^*} \tag{22}$$

where $\Gamma = (2\bar{v}_{Na^+} - \bar{v}_{Ca^{2+}})/\bar{v}_s$. This equation can be solved to obtain y*, which is again different from Donnan's criteria unless the left-hand side is equal to unity.

Next, we consider another case [Fig.5(a)]: a weak monoacid HB (such as acrylic acid) in compartment 1 with an equilibrium constant k$_{HB}$, neither HB nor B$^-$ can go through the membrane. We assume the other side a large reservoir with pure water only. We do not have ion conservation conditions as in the previous example because hydronium ion H$_3$O$^+$ and hydroxide ions OH$^-$ concentrations in the two compartments can adjust themselves via water autoionization reaction. This case is of interests, for example, for studying the ionic hydrogel expansion [Brannon-Peppas and Peppas, 1991]. Assuming a portion of HB is ionized, generating $B^-$ ions with a concentration y in the compartment 1. The corresponding $H_3O^+$ and $OH^-$ in compartment 1 will adjust to the new equilibrium, as shown in Fig. 5(b). The two ionization reactions are $HB + H_2O - H_3O^+ + B^-$, and $2H_2O - H_3O^+ + OH^-$. Since only $H^+$ and $OH^-$ can go through the membrane, we have

$$\bar{v}_{H_3O^+}\Pi + F\varphi + RT\ln\frac{y^*+z^*}{10^{-7}} = 0 \tag{23}$$

$$\bar{v}_{OH^-}\Pi - F\varphi + RT\ln\frac{z^*}{10^{-7}} = 0 \tag{24}$$

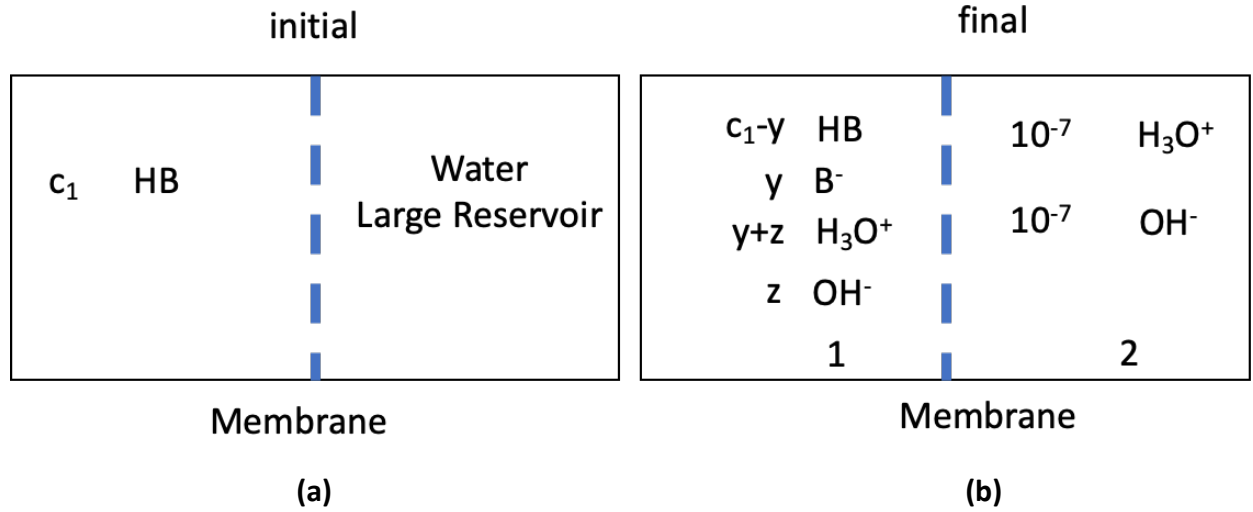

**Figure 5 Membrane equilibrium including water autoionization.** Weak acid HB in compartment 1, and water in compartment 2 (a) before and (b) after equilibrium. The compartment 2 is assumed to be large such that the concentration of the hydronium and hydroxide ions concentration do not change.

Adding the above two equations leads to

$$(\bar{v}_{H_3O^+} + \bar{v}_{OH^-})\Pi = -RT\ln\frac{(y^*+z^*)z^*}{10^{-14}} \tag{25}$$

Normally, we say that the water equilibrium constant $k_w=10^{-14}$ does not change, which means $c_{H^+,1}c_{OH^-,1} = y(y+z) = 10^{-14}$, and hence the osmotic pressure is zero. However, this is in conflict with the balance of the chemical potentials of $H_2O$ across the membrane, which leads to

$$\bar{v}_{H_2O}\Pi = -RT\ln\frac{\chi_{H_2O,1}}{\chi_{H_2O,2}} = -RT\ln\frac{1-(c_1^*+y^*+2z^*)}{1-2\times10^{-7}} \tag{26}$$

This apparent conflict is because we neglected the concentration change of water in the definition of the equilibrium constant. Strictly speaking, the equilibrium constant of water in the two compartments are

$$k'_{w,1} = \frac{c_{H_3O^+} \times c_{OH^-,1}}{c_{H_2O,1}^2} = \frac{(y+z)z}{[c_o-(c_1+y+2z)]^2} = \frac{k_{w,1}}{[c_o-(c_1+y+2z)]^2} \quad (27)$$

$$k'_{w,2} = \frac{c_{H_3O^+} \times c_{OH^-,1}}{c_{H_2O,1}^2} = \frac{10^{-14}}{[c_o-2\times10^{-7}c_o]^2} = \frac{k_{w,2}}{[c_o-2\times10^{-7}c_o]^2} \quad (28)$$

where we used superscript " $'$ " to denote the rigorous definition of the equilibrium constant including the molar fraction change of water accompanying the concentration changes of hydronium and hydroxide ions, and $k_{w,1}$ and $k_{w,2}$ are the conventional definition of the equilibrium constants.

Taking the ratio of Eqs. (25) and (26) leads to

$$\frac{(y+z)z}{10^{-14}} = \left[\frac{c_o-(c_1+y+2z)}{c_o-2\times10^{-7}c_o}\right]^{\Gamma} \quad (29)$$

where $\Gamma = (\bar{v}_{H_3O^+} + \bar{v}_{OH^-})/\bar{v}_{H2O}$. Although it is natural to assume that the sum of molar volumes of the hydronium and hydroxide ions equals twice the water molar volume, i.e., $\Gamma=2$, in reality, the molar volumes of the hydrogen ion and the hydroxide ion are not equaling to each. In fact, they can be negative [Marcus, 2012]. From Eqs. (27) and (28), we get the relationship between the water equilibrium constants between the two regions,

$$\frac{k'_{w,1}}{k'_{w,2}} = \left(\frac{c_o-(c_1+y+2z)}{c_o-2\times10^{-7}}\right)^{\Gamma-2} \quad (30)$$

We see that unless $\Gamma=2$, $k'_{w,1} \neq k'_{w,2}$. This difference in the equilibrium constants arise from the pressure dependence of the equilibrium constant. In fact, using Eq. (2), we can show

$$k'_w = \frac{c_{H_3O^+} \times c_{OH^-}}{c_{H_2O}^2} = k'^o_w exp\left(\frac{(\Gamma-2)\bar{v}_{H2O}(P-P_o)}{RT}\right) \quad (31)$$

where $k'^o_w$ is the equilibrium constant at the standard state. Applying the above relationship for the two sides of the membrane and using Eq. (26) leads to Eq. (30), which supports our previous statement that the difference in the equilibrium constant as in Eq. (30) is due to its pressure dependence.

The same argument applies to the equilibrium constant for BH, which leads to

$$\frac{k'_{BH,1}}{k'_{BH,o}} = \left(\frac{c_o-(c_1+y+2z)}{c_o}\right)^{\Gamma_{BH}-1} \quad (32)$$

where $\Gamma_{BH} = (\bar{v}_{H_3O^+} + \bar{v}_{B^-} - \bar{v}_{HB})/\bar{v}_{H_2O}$ , $k'_{BH,o}$ is the equilibrium constant at the standard condition, which again carries the concentration of the solvent as Eq. (27) for water. With $k'_{BH,1}$ known, the other needed equation to solve for y and z is

$$\frac{(y+z)y}{(c_1-y)} = k_{BH} \left(\frac{c_o-(c_1+y+2z)}{c_o}\right)^{\Gamma_{BH}} \tag{33}$$

In solving the above equations, one typically finds z is very small and can be neglected, and hence Eq. (33) alone can be used to find y, which is consistent with the conventional treatment in chemical equilibrium neglecting the water autoionization. Consider the special case y is much smaller than $c_1$ and $\Gamma_{BH} = 1$ i.e., then,

$$y = \left[k_{BH} c_1 \left(\frac{c_o-c_1}{c_o}\right)\right]^{1/2} \tag{34}$$

We can use the above y to compute contributions to the classical Donnan potential from $H_3O^+$ ion

$$\varphi_c = -\frac{RT}{2F} \ln\left[\frac{k_{BH}}{k_w} c_1 \left(\frac{c_o-c_1}{c_o}\right)\right] \tag{35}$$

The contributions to the Donnan potential from the osmotic pressure can be calculated from Eqs. (23) and (26)

$$\varphi_\Pi \approx -\Gamma_{H_3O^+} \frac{RT}{F} c_1^* \tag{36}$$

where $\Gamma_{H_3O^+} = \bar{v}_{H_3O^+}/\bar{v}_{H_2O}$ . Comparing Eqs. (35) and (36), we see that the two contributions are comparable when $\frac{k_{BH}}{k_w} \sim 1/c_1$.

From the above examples, we see that the osmotic contribution to the membrane potential is typically of the order of

$$\varphi_\Pi \sim -\frac{RT}{F}(c_1^* - c_2^*) \tag{37}$$

where $c_1^*$ and $c_2^*$ are the mole fraction of the impermeable and mobile species, respectively. The ion concentration difference contribution to the Donnan potential is

$$\varphi_c \sim -\frac{RT}{F} \ln \frac{x_1^+}{x_2^+} \tag{38}$$

From the above relation, we can see that the osmotic pressure contribution is negligible when

$$(c_1^* - c_2^*) \ll \ln \frac{x_1^+}{x_2^+} \tag{39}$$

When the concentration difference of the ions in the two compartments is large, the osmotic pressure contribution to Donnan potential is not negligible.

**Potential Arising from Osmosis**

The above analysis leads to a new prediction: even when the impermeable species is non-ionizable, there could exist a potential across the membrane. To see this, we assume now that HB in Fig.5 is completely non-ionizable, i.e, y=0. Subtracting Eq. (23) from (24), we get

$$\varphi = -\frac{1}{2F}(\bar{v}_{H_3O^+} - \bar{v}_{OH^-})\Pi \approx -\frac{RT}{2F}\frac{(\bar{v}_{H_3O^+} - \bar{v}_{OH^-})}{\bar{v}_{H2O}}c_1^* \qquad (40)$$

where the last step is obtained by replacing the osmotic pressure by the van't Hoff formula, assuming that $c_1$ is the dominant concentration, which is justifiable since we can solve for z from Eq. (21) approximately by neglecting both $10^{-7}$ and z term in the right-hand side:

$$z = \sqrt{k_w}(1 - c_1^*)^{\Gamma/2} \qquad (41)$$

which is a small number. From Eq. (40), we can see that as long as the molar volume of the hydronium ions do not equal to that of the hydroxide ions, a potential difference will exist across the membrane. This potential difference balances the ion motion created by the osmotic pressure. According to Marcus [2012], at 25 °C, $\bar{v}_{H^+} = -5.1$ cm$^{-3}$/mole, $\bar{v}_{OH^-} = 1.2$ cm$^{-3}$/mole. For water molecules, $\bar{v}_{H2O} = 18$ cm$^3$/mole. We take $\bar{v}_{H_3O^+} = \bar{v}_{H^+} + \bar{v}_{H2O} = 12.9$ cm$^{-3}$/mole. The above relation leads to $\varphi = -8.4c_1^*$ mV. This membrane potential should be readily measurable.

**Discussion**

Our analysis shows that the key to the coupling is the solvent equilibrium. Donann's original experiments intentionally avoided water osmosis. For example, Donnan and Garner [1919] wrote "When solutions of different concentrations were employed on the two sides of the membrane, osmosis of water was prevented by the addition to the solution of the requisite amount of sucrose." However, they also said the effect was small, clearly due to the dilute nature of the solutions they considered. Our analysis above shows that the coupling between the ion concentration with osmotic pressure becomes important when the mole fraction of the solvent, or the ratio of the mole fraction between the two solutions, deviates significantly from unity, i.e., when the concentrations of the solutes are large [one can easily see this from Eq. (10)]. Since water content in cells are typically around 70%, this coupling effect might be important. In gels and membranes, batteries and supercapacitors, solvent content could also be much less than unity, signaling the potential importance of considering the coupling effect discussed.

In fact, this coupling could be important even in dilute solutions near a solid surface, a charged polymer chain, or an ion, around which the local ion concentrations can be very high. Consider, for example, the electrical double layer around an ion or a flat surface, the counterion concentrations near the surface can be extremely high. In fact, predictions of counterion mole fraction from the Poisson-Boltzmann equation can exceed unity [Bazant et al., 2009]. This difficult was circumvented with the existence of a Stern layer [Stern, 1924] or Manning condensation layer [Manning, 1969]. The fundamental reason is however because the Poisson-Boltzmann equation ignored completely the solvent equilibrium. Different modifications of the Poisson-Boltzmann equations had been proposed, most of them did not include the solvent effect neither, except Freise [1952], who started from the differential form of Eq. (2) and included the solvent equilibrium. Results from Freise' equations for the electrical double layer are consistent with the coupled equilibrium criteria discussed here, and naturally include the ion crowding effect. In fact, the Poisson-Bolzmann equation is apparently consistent with the classical Donnan potential [Ohshima and Ohki, 1985] since both neglected the solvent equilibrium. So, the equilibrium of an electrical double layer can also be thought as the Donnan equilibrium. Although never tried, Freise's approach should be able to explain the minimum of the activity coefficient often observed in concentrated electrolytes that cannot be predicted by the Debye-Huckle theory [Debye and Huckle, 1923], as one can infer from the success of Egan and Wicke [1954] who modified the Boltzmann distribution to include the crowding effect. Such coupling effect could also be important for colloids stabilized by the competition between the double layer and the Casimir force [Derjaguin and Landau, 1993; Verwey and Overbeek, 1948]. Interesting, even Overbeek, one of the developers of the DLVO theory for the colloid stability, did not recognize the importance of the pressure term when he considered the local high concentration of counterions [Overbeek, 1956].

The Nernst-Planck equation, coupled to the Poisson equation (also called Poisson-Nernst-Planck equation), are often used in simulating the transport of ions [Probstein, 1994; and Bazant et al., 2004]. At steady-state and in the absence of transport, these sets of equations can lead to the Poisson-Boltzmann equation. This implies an inherent inconsistency of the Nernst-Planck equation with the coupled equilibrium criteria discussed here. This inconsistency is because the Nernst-Planck equation also neglected the coupling between the solvent and ions in the transport via the pressure term in Eq.(2). Schlogl [1966] started from Eq. (2) and derived a set of equations called extended Nernst-Planck equations, which included such coupling (Schlogl was aware of the Freise work). Schlogl's extended Nernst-Planck equations are used to model membrane transport [Mohammad et al., 2015]. Hence, the coupled Donnan equilibrium criteria discussed in this paper are inherently consistent with the work of Freise and Schlogl.

**Concluding Remarks**

To summarize, although Donnan membrane equilibrium criteria are well established, they neglected the coupling between the osmotic pressure and the ion concentration. A proper treatment of the membrane equilibria requires the simultaneous consideration of the balance of the chemical potentials among the ions and between the solvents in the membrane-separated regions, which couples the membrane potential, the osmotic pressure and the concentrations. Due to this coupling, the membrane potential not only includes the traditional Donnan term, but also an additional term due to osmotic pressure. This coupling leads to our prediction that a membrane potential should exist even if the immobile species is not charged.

The modified criteria for Donnan equilibrium discussed here had been established before [Morales and Shock, 1941; Scotto et al., 2016], but these studies did not draw attention as is evident from the fact that they were cited only 1 and 2 times, respectively. Freise's work [1952] was cited 105 time (citation picked up since 2005 thanks to the careful review by Bazant et al. [2004]) and Schlogl's work [Schlogl, 1966] 26 times, in stark contrasts to the tremendous amount of work based on the Poisson-Boltzmann equation and Nernst-Planck equation. We have shown the inherent consistency of the coupled treatment of the Donnnan equilibrium with the Freise and the Schlogl approaches. The essence of the coupling is that we need to consider the coupling between the solvent equilibrium (or its transport) with the ions via pressure common to them all. Due to the wide range of applications of Donnan equilibrium, and the intrinsic connection of the coupled treatment with that of Freise and Schlogl, I believe that are urgent need to validate this coupling experimentally, to apply it to re-assess the established problems and to solve new problems. Some of the outstanding questions are

- How can we experimentally demonstrate the importance of the coupling discussed? One such experiment, for example, is to validate the prediction we made here on the existence of a membrane potential even though the immobile ions are not charged.

- How does the discussed coupling impact classical problems such as the activity coefficients of electrolytes, the DLVO theory, since the electrical double layer underlying these theories could be thought as in generalized Donnan equilibrium state, despite the absence of membranes.

- The Nernst equation that forms the basis of electrochemistry is a special case of the Donnan equation. Similarly, it does not include the coupling effect discussed here. How does the coupling discussed here impact the voltage of electrochemical systems such as batteries?

- How does the coupling effects discussed here impact problems in biology, batteries and supercapacitors, hydrogels, membranes for separation, and electrokinetic and electro-osmotic flows?

Furthermore, the discussion here focused on constant temperature system. The same strategy could be extended to include effects of inhomogeneous temperature. Such consideration might be important for emerging electrochemical waste heat recovery technologies [Han et al., 2020].

**Acknowledgements:** In 2016, I had the honor of receiving the prestigious Eringen medal. In thinking of what I can contribute to celebrate the centennial anniversary of Professor Eringen, I feel that some discussion on the Donnan equilibrium established also ~100 years ago might be appropriate. I recently got interested in this topic and found that some aspects had been neglected for since Donnan's time. Although I had not worked on the subject before, my curiosity led me to develop the work independently before I found the work of Scotto et al., and later Freise and Schlogl, and Morales and Shock. I would like to thank Dr. Tom Dursch for his encouragement and also pointing out the work of Overbeek, and Professors James Lee, Jiaoyan Li, and Leyu Wang for their kind invitation. I would particularly like to thank MIT for its support in a difficult time of my life that enabled me to explore the topics discussed in this work.